# MOVING-BASE GRADIOMETRY WITHOUT GRADIOMETERS: BACK TO THE FUTURE

**Alexey V. Veryaskin**
*Trinity Research Labs*
*The Australian International Gravitational Research Centre (AIGRC)*
*University of Western Australia*
*35 Stirling Highway, Nedlands, WA6009*
*alexey.veryaskin@uwa.edu.au*

## SUMMARY

Moving-base Gravity Gradiometry has emerged from the need to cancel out the effect of kinematic accelerations on gravity measurements in motion. According to the Einstein's Equivalence Principle, a gravimeter mounted on a moving platform cannot distinguish between the force of gravity and the inertial force acting on it within the same frequency bandwidth at the same time. Gravity Gradiometers measure the first spatial derivatives of the combined gravitational and inertial forces. As kinematic accelerations do not possess any spatial gradients, their influence is cancelled out making the moving-base gravitational measurements possible. Another possibility of moving-base gravitational measurements without the use of real gravity gradiometers is discussed here. One can show that the same approach is directly applicable to moving-base magnetic gradient measurements as well. This possibility opens a door for the development of ultra-miniature, UAV deployable and cost effective moving-base geophysical exploration systems.

## HISTORICAL BACKGROUND

In 2002, Alan H. Zorn [1] published the following statement: "It is demonstrated that a complete solution to real-time navigation and gravity gradient determination can be performed simultaneously and unambiguously using all-accelerometer inertial measurements only. Although the separation of gravity from kinematic motion appears to violate Einstein's principle of equivalence, this is not the case". He also published an equation, which links absolute gravity gradients with full time derivative of the gravitational acceleration if measured on a moving platform:

$$\frac{d}{dt}\vec{g} = (\vec{V}\vec{\nabla})\vec{g} = \left(V_x\frac{\partial}{\partial x} + V_y\frac{\partial}{\partial y} + V_z\frac{\partial}{\partial z}\right)\vec{g}, \quad (1)$$

where

$$\vec{\nabla} = \left(\frac{\partial}{\partial x}, \frac{\partial}{\partial y}, \frac{\partial}{\partial z}\right),$$

$\vec{V} = (V_x, V_y, V_z)$ is the velocity vector of the moving platform with respect to Earth's surface and $\vec{g}$ is the gravitational acceleration vector.

Equation (1) follows directly from a well-known equality [2]

$$\frac{d}{dt}f(\vec{r},t) = \frac{\partial}{\partial t}f(\vec{r},t) + (\vec{V}\vec{\nabla})f(\vec{r},t) \quad (2)$$

An observer measuring the full time derivative of a physical quantity $f$, possessing spatial gradients, will see not only its temporal variations in the moving reference frame, but also the rate at which it varies from one spatial point to another. The latter one is proportional to the speed of motion.

Unlike the method discussed in [1], I shall explore below another alternative for moving-base gravity gradient measurements, which is based directly on Eq.2. The same approach can be used for moving-base magnetic gradient measurements as well.

## METHOD AND RESULTS

Let us consider a conventional precision accelerometer mounted on a platform moving in, say, Y direction as depicted in Fig.1 below. The accelerometer can be treated as a mechanical oscillator having a proof mass m with an effective spring constant, and a well-defined sensitivity axis. Let's assume for a moment that the sensitivity axis is aligned along the X-axis and the platform does not experience any angular motion. In the real world, the latter does not hold and the angular motion must be taken into account (see below). In the moving reference frame, the accelerometer's equation of motion is as follows

$$\frac{d^2}{dt^2}\Delta X + \frac{2}{\tau^*}\frac{d}{dt}\Delta X + \Omega_{eff}^2\Delta X = \Delta g_x - \tilde{a}_x + \tilde{N} \quad (3)$$

Within a frequency band that is much lower than the effective resonant frequency of the accelerometer ($\Omega_{eff}$) the displacement of the proof mass, as a function of applied kinematic acceleration $a_x$ and variation of the gravitational acceleration $\Delta g_x$, is expressed as follows:

$$\Delta X = \frac{\Delta g_x - \tilde{a}_x + N}{\Omega_{eff}^2} \quad (4)$$

Here $\Delta x$ is the accelerometer's proof mass relative mechanical displacement with respect to its reference position. For example, in MEMS based accelerometers with capacitive sensing, $\Delta x$ is the gap between the capacitive sensing plates.

$N$ is the accelerometer intrinsic mechanical displacement noise, which may include not only white noise but also a 1/f term.

For the sake of simplicity, I shall not discuss here any non-linear effects, which typically are in place in all kinds of mechanical accelerometers. These effects are well known and can be cancelled out by using feedback locked-loop sensing techniques.

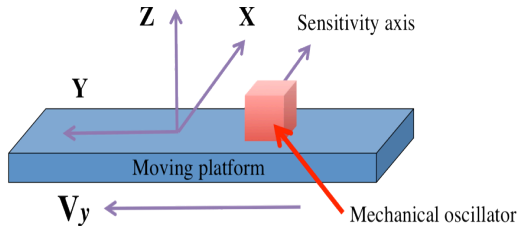

**Figure 1. An accelerometer mounted on a moving platform is treated as a linear mechanical oscillator with its sensitivity axis aligned perpendicular to the direction of motion. This choice is not a necessary one and is only used as the simplest case to illustrate the non-conventional approach to the moving-base gravity gradient measurements.**

Instead of traditional mechanical displacement measurements, the accelerometer's velocity measurement yields

$$\frac{d}{dt}\Delta X = v_x = \frac{1}{\Omega_{eff}^2}\left(\frac{d}{dt}g_x - \frac{\partial}{\partial t}a_x + \frac{\partial}{\partial t}N\right) = \\ = \frac{1}{\Omega_{eff}^2}\left(V_y\Gamma_{xy} + \frac{\partial}{\partial t}g_x - \frac{\partial}{\partial t}a_x + \frac{\partial}{\partial t}N\right) \quad (5)$$

Then the $\Gamma_{xy}$ component of the gravitational gradient tensor can be recovered as follows

$$\Gamma_{xy} = \Omega_{eff}^2\left(\frac{v_x}{V_y}\right) - \frac{1}{V_y}\frac{\partial}{\partial t}g_x + \frac{1}{V_y}\frac{\partial}{\partial t}a_x - \frac{1}{V_y}\frac{\partial}{\partial t}N \quad (6)$$

The term proportional to the temporal variations (tidal variations for example) of the gravitational acceleration in the right side of Eq.6 can be ignored for fast moving platforms, even if one aims at achieving 1 Eotvos measurement noise floor.

It is interesting to notice that the gravity gradient component in Eq.6 is expressed in absolutes units. Similarly, it is well known that if a relative gravimeter is moved from one spatial position to another, the result of the measurement is proportional to an absolute value of a gravity gradient component.

It is also interesting to notice that the gravity gradient component in Eq.6 is relevant to the Earth gravitational field only. The platform's local gravity gradients that must be picked up if the measurements are done by a conventional gravity gradiometer, are not measured as the corresponding gravitational field source moves with the same speed.

Further discussions may suffer a criticism from a common view that one must set up a facility for independent measurements of the kinematic accelerations without measuring the gravitational one. The latter one is not possible according to the Einstein's Equivalence Principle. This is only true if there is nothing outside of the moving platform that can be used as a stationary reference frame.

One of the possibilities is to use Global Positioning System (GPS) for the independent velocity and acceleration measurements. The platform moves relatively to a GPS and, generally speaking, its velocity and acceleration can be determined without using conventional accelerometers. The latter ones, indeed, cannot distinguish between the inertial and gravitational forces.

Unfortunately, the accuracy of measuring velocity and its time derivatives by means of GPS is not quite adequate to the demand of having 1 Eotvos measurement error for moving-base gravity gradient measurements.

**LORENTZ-FORCE ACCELEROMETER**

Another physical reference frame, which is stationary (inertial) with respect to a moving platform, is the Earth's magnetic field.

If an electric charge is used instead of a proof mass, one can measure the platform motion relatively to the Earth's surface by the use of the Lorentz force [3] acting upon the charge. The Lorentz force is insensitive to the force of gravity. So, in theory, a Lorentz-force based accelerometer [4] can be used as the secondary accelerometer mounted on the moving platform and measuring the kinematic accelerations for the purpose described above.

It should be underlined that the Lorentz-force accelerometer concept is just a theoretical possibility at this stage. However, the first experimental results that support this possibility have been reported [4].

**SYSTEM REQUIREMENTS**

The next logical step is to estimate the limitations coming from the other terms in Eq.6, such as the first time derivative of the kinematic acceleration experienced by the measuring accelerometer in the moving reference frame. Also, one can estimate the acceptable error in measuring the platform's speed and the fundamental thermal noise limit inherent to all measurement techniques.

The requirement to have 1 Eotvos ($10^{-9}$ $1/sec^2$) measurement error per 1 sec measurement time is chosen to be a benchmark, according to the common dream of the geophysical exploration community.

It is assumed that the averaged platform speed is 50 m/sec, which is typical for the light and medium class of airborne platforms. I shall omit here the mathematical exercise, which has led to the results shown below.

The fundamental thermal noise limit (one sigma)

$$\sigma_\Gamma^{(T)} = \frac{2\pi}{V\tau}\sqrt{\frac{k_B T}{m\tau^*\tau}} \cong 0.3E \quad (7)$$

Here V=50 m/sec, T=300 K is the room temperature, m=1 gram is the proof mass, $\tau^*$ is the mechanical oscillator relaxation time and τ is the measurement time. It is assumed that they are both equal to 1 sec.

Eq.7 is an interesting one. The only difference between the standard thermal noise floor that is inherent to any kind of proof mass based gravity gradiometers and the Eq.7, is that the latter one does not contain a base line. It contains a product of Vτ, which has the same dimension. A single accelerometer does not possess a base line. The latter one is an integral part of only differencing types of gravity gradiometers. By increasing the platform speed, the equivalent "base line" can be made much longer compared to that of typical size gravity gradiometers. "The faster flying the better" is the characterisation of this new approach, which is entirely opposite to "the slower flying the better" one which has been exercised in relation to the only flying rotating the "Falcon" and the FTG technologies [5].

The kinematic acceleration $a_x$ in Eq.6 must be measured with the maximum error of

$$\Delta \tilde{a}_x < \frac{V}{2\pi \Delta f} 10^{-9} \cong 4 \; 10^{-7} \; \frac{m}{\sec^2} \tag{8}$$

Otherwise, the platform must be stabilised at the same level by using standard stable table facilities. Again, here V=50 m/sec and $\Delta f \leq 0.02$ Hz is the useful frequency bandwidth of the gravity gradient data.

The most controversial requirement is that the kinematic acceleration must be measured by a non-conventional accelerometer as per said above.

From Eq.6, one can estimate the error the platform speed must be measured with (or equally kept stable)

$$\Delta V \leq \frac{V}{\Gamma_{\max}} 10^{-9} \cong 10^{-2} \; \frac{m}{\sec} \tag{9}$$

where $\Gamma_{\max} \sim 3000$ Eotvos is the maximum gravity gradient on the Earth surface and V=50 m/sec. This requirement is within the reach of modern GPS capabilities.

According to Eq.6, the main accelerometer performance, that is required in order to comply with the 1 Eotvos benchmark noise floor, is as follows

$$\Delta v \leq \frac{V}{\Omega_{eff}^2} 10^{-9} \; \frac{m}{\sec} \tag{10}$$

within the $\Delta f \sim 0.02$ Hz frequency band.

**PLATFORM'S ANGULAR MOTION EFFECTS**

It is well known that linear mechanical oscillators are not sensitive to any kind of angular motion effects in the rotating reference frame except that their mechanical susceptibilities are weakly coupled to angular velocities (see Eq.11 below). However, variations in the orientation of the main accelerometer sensitivity axis with respect to the stationary reference frame must be taken into account.

Assuming that the moving platform is angularly stabilised anyway and that its residual angular motion is small enough, the leading corrections to Eq.6 are as follows:

$$\Gamma_{xy} = \Omega_{eff}^2 \left( \frac{v_x}{V_y} \right) - \frac{1}{V_y}(-g_y \Omega_z + g_z \Omega_y) + \frac{1}{V_y} \frac{\partial}{\partial t} a_x - \frac{1}{V_y} \frac{\partial}{\partial t} N$$

$$\Omega_{eff}^2 = \omega_m^2 - \Omega_y^2 - \Omega_z^2 \tag{11}$$

where $\Omega_x$, $\Omega_y$, and $\Omega_z$ are the platform's angular velocity vector components correspondingly.

It is quite straightforward to determine the angular stabilisation requirements that any gravity gradiometer technology needs to provide anyway. The major error term comes from the large vertical acceleration of gravity (~10 m/sec$^2$) being projected onto the accelerometer sensitivity axis (X-projection in our case). The corresponding gravity gradient error is as follows

$$\Delta \Gamma_{xy} = \frac{1}{V}(g_z \Delta \Omega_y + \Delta g_z (\Omega_y)_{\max}) \tag{12}$$

The first error term is caused by the error in measuring the platform's angular rates and using the data to actively stabilise one. The second error term is the error caused by the accuracy of the Earth Gravitational Model (EGM). It is proportional to the residual angular velocity that platform can sustain in order to provide the 1 Eotvos benchmark measurements. Eq.12 yields

$$\Delta \Omega < \frac{V}{g_z} 10^{-9} \cong 5 \; 10^{-9} \; \frac{rad}{\sec} \tag{13}$$

The angular stabilization requirement, as per Eq.13, can be fulfilled only within the useful frequency band, which is typically below 0.02 Hz. Fiber-Optic Gyros (FOG) are good candidates for achieving the goal as they have proven to provide a white noise level of $10^{-4}$ deg/√hour or $3 \; 10^{-8}$ (rad/sec)/√Hz [7].

The statement above applies only if a post processing stage is an integral and the final part of the measurements. There is no room for real-time data processing as yet.

The total angular motion coupled dynamic range for the real moving-base environment can be determined by a pre-stabilisation stage by using a compact commercially available stable platform. The following figures are used to work out the second error term in Eq.12, as per, as an example, *Leica PAV100* stable table technical specifications [8]:

$$\Omega_{\max} = 2\pi (\Delta f_{realtime}) \Delta \varphi_{\max}, \; \Delta \varphi_{\max} = 3 \; 10^{-4} \; rad$$

$$\Delta f_{realtime} = 0.5 \; Hz, \; V = 50 \; \frac{m}{\sec} \tag{14}$$

Where $\Delta f_{realtime}$ is the real time measurement frequency bandwidth, which corresponds to 1 sec measurement time. This yields

$$\Delta g_z < \frac{V}{(\Omega)_{\max}} 10^{-9} \cong 5 \; mGal \tag{15}$$

All error terms shown above are just rough estimates and a more detailed error budget analysis shall be provided elsewhere.

**GRAVITY GRADIOMETRY vs MAGNETIC GRADIOMETRY**

In the 1950s, Warren E. Wickerham [9] was the first to suggest that a linear sweep rate of a total field magnetometer, along a line of survey, produces a signature similar to that of a conventional magnetic gradiometer.

Another example of a moving-base magnetic gradiometer concept, based on Eq.2, is shown in Fig.2 below. It is easy to prove that the total EMF generated in a moving (in the same Y-direction) induction coil results in the following voltage output $V_{out}$ fed forward by an adjacent opamp.

$$\begin{aligned} V_{out} &= S_{eff} \frac{d}{dt} B_y + \sqrt{4 k_B T R \Delta f} + \tilde{N}_A \\ &= S_{eff}(V_y B_{yy} + \frac{\partial}{\partial t} B_y - B_{x0}\Omega_z + B_{z0}\Omega_x) + \\ &\quad + \sqrt{4 k_B T R \Delta f} + \tilde{N}_A \end{aligned} \quad (16)$$

where $B_{yy}$ is the magnetic gradient component, which, then, can be recovered as follows

$$B_{yy} = \frac{V_{out}}{V S_{eff}} - \frac{1}{V}(\frac{\partial}{\partial t} B_y - B_{x0}\Omega_z + B_{z0}\Omega_x) - \frac{1}{V S_{eff}}(\sqrt{4 k_B T R \Delta f} + \tilde{N}_A) \quad (17)$$

where V is the platform's speed in m/sec, $S_{eff}$ is the coil's effective area in $m^2$, $N_A$ is the opamp input noise in volt and $\Delta f$ is the useful measurement bandwidth.

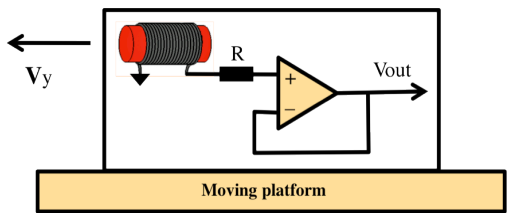

**Figure 2.** The simplest realisation of a moving-base magnetic gradiometer concept based on Eq.2. This block diagram however is not a practical one as typically all opamps exhibit large 1/f noise below 1 Hz in the spectral domain.

One can argue that sub-pT/m at 1 sec measurement time is feasible to achieve. The main problem for this type of measurements is the temporal variations of the Earth magnetic field (magnetotelluric) and local mount-based quiasi-static magnetic field disturbances. The system would require a fusion of a precision vector magnetometer and a commercial grade Gyros. A strap-down deployment is the preferable choice as there must be no any relative motion between the coil (see Fig.2 above) and other metal parts (if any) of the platform mount.

A practical realisation of the moving-base magnetic gradiometer concept based on an effective modulation-demodulation technique and a detailed error budget analysis shall be reported in due course.

**DISCUSSIONS**

It seems that the major problem that prevents the geophysical exploration community from hearing good news about deploying a new moving-base gravity gradiometer is the size of the stable table that is required to accommodate one. Even for the MEMS-based gravity gradiometers [6], there must be a reasonable base line, which determines the sensitivity of the gradiometer and its size. The larger and more cumbersome the stable platform is, the more difficult to actively control its rotational and vibrational degrees of freedom.

As the single accelerometer approach does not require any base line, it can lead to an extremely compact system fusion. The system can be miniaturised to the extend that it can fit into the UAV-based moving environment. Currently, there are not any moving-base gravity gradiometer designs for the terrestrial applications that can be made UAV deployable.

Modular (easy accessible) system fusion based on available off-the-shelf advanced nano-positioning stages with feedback control and the progressing MEMS technology would provide the necessary and cost-effective replaceable componentry.

The conventional gradiometers are still needed, as the approach discussed above does not work while stationary. A fast moving platform is the necessary integral part in the whole equation.

**ACKNOWLEDGMENTS**

I would like to thank Prof David Blair and Prof Ju Li of the AIGRC for their persistent support for my works in the area of gravity, magnetic and EM gradiometry. I would also like to thank Prof Ho Jang Paik and Dr Andrew Sunderland for useful discussions in relation to the new approach to moving-base gravity and magnetic gradiometry. I do appreciate support from Prof Jun Luo and Prof Liang Cheng Tu of the Huazhong University of Science and Technology (The Key Laboratory of Fundamental Physics Quantities Measurements, Wuhan, P.R. China) in participating jointly in the "111 Project", which is partially aimed at some non-conventional geophysical exploration technologies.

**REFERENCES**

[1] Zorn, A.H., 2002, A Merging of System Technologies: All-Accelerometer Inertial Navigation and Gravity Gradiometry: IEEE Position Location and Navigation Symposium, Palm Springs, California, April 15-18.

[2] Landau, L.D. and Lifshitz, E.M., 1971, The Classical Theory of Fields: Volume 2 of the Course of Theoretical Physics, Pergamon Press, p.47.

[3] Jackson, J.D., 1999, Classical Electrodynamics (3rd ed.): New York, Wiley, p.191.

[4] Veryaskin, A.V., 2014, A talk presented at the Huazhong University of Science and Technology: The Key Laboratory for Fundamental Physics Quantities Measurements, Wuhan, P.R. China (unpublished).

[5] Lee, J.B. et al, 2001, High resolution gravity surveys from a fixed wing aircraft: Geoscience and Remote Sensing Symposium: IEEE 2001 International Volume 3, 1327-1331.

[6] Cuperus, R. et al, 2009, Gravity Gradiometer System for Earth Exploration: Micronano Conference, 5-6 Nov, Delft, The Netherlands.

[7] Lefevre, H.C., (2012), The fiber-optic gyroscope: Achievements and perspective, Gyroscopy and Navigation, v.3(4), 223-226.

[8] http://www.leicageosystems.com/en/LeicaPAV100_103713.htm

[9] Wickerham, W.E., (1954), The gulf airborne magnetic gradiometer: Geophysics, v.19(1), 116-123.